\documentclass{aa}  
\usepackage{graphicx}
\usepackage{color}
\usepackage{ulem}
\usepackage{latexsym}
\usepackage{natbib}
\bibpunct{(}{)}{;}{a}{}{,} 
\usepackage{txfonts}
\newcommand{\msun}{\mbox{$M_{\odot}$ }}
\newcommand{\e}[1]{\cdot 10^{#1}}

\begin{document}
	\title{Variations in Integrated Galactic Initial Mass Functions due to Sampling Method and Cluster Mass Function}
	\titlerunning{Variations in IGIMFs}

   \author{M.R. Haas
          \inst{1}
          \and
         P. Anders \inst{2} 
          }

   \offprints{M.R. Haas, \email{haas@strw.leidenuniv.nl}}

   \institute{	Leiden Observatory, Leiden University, 	P.O. Box 9513, NL-2300 Leiden, The Netherlands
        \and	 
		Astronomical Institute, Utrecht University, Princetonplein 5, NL-3584 CC Utrecht, The Netherlands
                   }

   \date{Received ; accepted }

 
  \abstract
	{Stars are thought to be formed predominantly in clusters.
The star clusters are formed according to a cluster initial mass
function (CMF) similar to the stellar initial mass function (IMF). 
Both the IMF and the CMF can be approximated by (broken)
power-laws, which favour low-mass objects.
The numerous low-mass clusters will lack high mass
stars, compared to the underlying IMF, since the most massive star
cannot be more massive than its host cluster.
If the integrated galactic initial mass function (IGIMF, i.e. the  total
stellar mass function of all stars in a galaxy) originates from stars
formed in star clusters, the IGIMF could be steeper than the IMF
in clusters.}  
	{We investigate how well constrained this steepening is and how it
depends on the choice of sampling method and CMF. We 
investigate the observability of the IGIMF effect in terms of galaxy 
photometry and metallicities.}   
	{We study various ways to sample the stellar IMF within star clusters
and build up the IGIMF from these clusters. We compare analytic sampling
to several implementations of random sampling of the IMF, and
different CMFs. We implement different IGIMFs into the \textsc{galev}
evolutionary synthesis package to obtain colours and metallicities for galaxies.}  
	{Choosing different ways of sampling the IMF results in
different IGIMFs.
Depending on the lower cluster mass limit and the slope of the cluster
mass function, the steepening varies between very strong and
negligible. We find the size of the effect is continuous as a function of the power-law slope of the CMF, if the CMF extends to
masses smaller than the maximum stellar mass.
The number of O-stars detected by GAIA will, if some
uncertain factors are better understood,  help in judging on the
importance of the IGIMF effect. The impact of different IGIMFs on
integrated galaxy photometry is small, within the intrinsic scatter of
observed galaxies. Observations of gas fractions
and metallicities could rule out at least the most
extreme sampling methods, if other sources of error are sufficiently 
understood. }  
	{As we still do not understand the details of star formation and
the sampling of the stellar IMF in clusters, one sampling method cannot
be favoured over another. Also, the CMF at very low cluster masses is
not well constrained observationally. These uncertainties therefore need
to be taken into account when using an IGIMF, with severe implications
for galaxy evolution models and interpretations of galaxy observations.}


   \maketitle
%

\section{Introduction}

A series of papers \citep[][the latter WK06 from now
on]{kroupaweidner03, weidnerkroupa04, weidnerkroupa05, weidnerkroupa06}
pointed out that the distribution of initial stellar masses in a galaxy
may significantly deviate from the initial mass function (IMF) the stars
have when they are born, if the vast majority of stars is born in
clusters. These clusters  follow a power-law mass function (the cluster mass
function, CMF), such that most stars form in low-mass clusters. In
low-mass clusters there is a deficiency of massive stars (e.g., as the
most massive star cannot exceed the total cluster mass), resulting in an
integrated galactic initial mass function (IGIMF) that is, at the high
mass end, steeper than the IMF.

The universality of the IMF is still an often debated topic. It is as yet not
clear whether the IMF in all Galactic star clusters is the same, whether or
not the field stars in the Milky Way follow the same mass distribution as
cluster stars, and whether the IMF in other galaxies is the same as here.
The IMF is shaped by the very complicated
processes which transform molecular cloud cores into stars, processes which
would be expected to be environmental-dependent. Therefore, a
non-universality of the IMF would intuitively be expected.

As the distribution of stellar masses has a profound impact on many aspects
of the evolution of galaxies, it is important to know to what extent the
IGIMF deviates from  the underlying stellar IMF (which is often used as IGIMF
in galaxy evolution studies) and how this affects galaxy properties.  For
example, the relation between star formation rate and H$\alpha$ luminosity is
shown to be steeper in galaxies with a very low star formation rate
\citep{skillman03}, which can be explained by having a steeper IGIMF for low
SFR galaxies \citep{pflammaltenburg07}, due to the preferential formation of
low-mass clusters. Also, the gradients in galactic disks of SFR and H$\alpha$
luminosity are different due to clustered star formation
\citep{pflammaltenburgkroupa08}. The supernova rate per unit stellar mass
formed and the  chemical enrichment history of a galaxy are influenced by the
IGIMF as shown by \citet{goodwinpagel05}. In a recent paper
\citet{hakobyan09} study the difference in rates of supernovae of type Ib/c
and type II and find that their results can be explained by having a steeper
IMF in the outskirts of galaxies than in their centers, which can be
explained by a different `IGIMF' in the outskirts of the galaxy as compared
to the inner regions due to a lower SFR in the outskirts.

Recently, \citet{recchi09} have investigated the [$\alpha/$Fe] versus 
velocity dispersion in early type galaxies and the rates of Supernovae
of  both Type II and Ia in several galaxy types, in the light of the
IGIMF  framework. They find that, if one assumes a constant star
formation rate  over a Hubble time, then for all but the irregular
galaxies these numbers agree well with the observed values. \citet{recchi09} explain this discrepancy by stating that for
irregular galaxies a constant SFR over the age of the Universe is not
likely to be a good approximation.

However, other studies (see, e.g., \citet{sandage86}) find approximately
constant SFR for late-type spiral galaxies (Sd/Irr), and declining SFRs
with time for earlier-type galaxies (where the decline time decreases
while going from Sc to E galaxies). For Sa-Sc galaxies, the SFR is
directly related to the available gas mass, resembling the
Kennicutt-Schmidt law (\citealt{kennicutt98}). Starbursts, superimposed
on any of the standard Hubble types, seem to be a common phenomenon.
They have the strongest impact on photometry and chemical enrichment for
late-type galaxies (which are typically of low mass) and major mergers
(due to the triggered extremely high SFRs). Such starbursts might be
interpreted as ``recently rising SFR'' as found by \citet{recchi09}.

WK06 test three different scenarios for sampling stellar masses in a
cluster. They conclude that `sorted sampling' (see 
Sect.~\ref{sec:sortedsampling})  best reproduces the observed relation
between maximum stellar mass in a cluster  and the cluster mass (but see
\citet{maschberger08} for a critical re-evaluation of this relation).
The amount of steepening of the IGIMF is found to depend  on the
sampling method and on the power-law index of the low-mass end of the
CMF.

For galaxies as a whole, the low-mass end of the CMF is not well
constrained. Even in the Milky Way we can only see low-mass star-forming
regions (few to few tens of solar masses) nearby, while for distant
galaxies such regions are too faint.

In this work we investigate the dependence of the IGIMF on

\begin{enumerate}

\item \textit{Sampling method:} stellar masses in clusters can be
sampled in different ways from the stellar IMF. We will show that the
specific sampling method is indeed important and different sampling
methods give different results, as was already shown by
WK06. We will extend their set of sampling methods.

\item \textit{Cluster mass function:} It is to be expected that the
effects on the IGIMF depend on the CMF. Sampling issues become more
important for low-mass clusters and therefore a lower minimum cluster
mass and/or a steeper CMF will result in a stronger steepening of the
IGIMF. We take observed CMFs for high mass clusters and extrapolate them
down to the masses of observed star forming regions in the solar
neighbourhood. We investigate the impact of different lower mass limits 
and power-law indices.

\end{enumerate}

In Sect. \ref{sec:galaxymodels} we implement some IGIMFs into the {\sc
GALEV}  galaxy evolution models \citep{bicker04,GALEV09}, which follow
the photometric and chemical history of idealized galaxy models
self-consistently. We will  investigate how properties like the
integrated broadband photometry in  several filters and total gas
metallicity are influenced by taking into  account sampling issues in
the IMF, and discuss observational needs to quantify the importance of
the IGIMF effects for galaxy evolution and observations of integrated
galaxy properties.

We will start by presenting the mass distributions of stars and clusters
that we use in Section~\ref{sec:method} and discuss our sampling methods,
including a  consistency test of the sampling methods in Section~\ref{sec:sampling}. The results for
the IGIMF are shown in Section~\ref{sec:results}, for several sampling
methods with a constant  cluster mass function and for one sampling
method with a variety of cluster mass  functions. In
Section~\ref{sec:Ostars} we calculate the number of O-stars that  will
be observed by GAIA, under various assumptions, and we compare the
results of  our IGIMFs with the work on single O-stars by
\citet{dewit04, dewit05}.  Section~\ref{sec:galaxymodels} describes the
galaxy evolution models and shows  results on the integrated photometry
and chemical enrichment of galaxies with  various IGIMFs. The
conclusions are presented in Section~\ref{sec:conclusions}.


\section{The underlying mass functions}
\label{sec:method}

In this section we will discuss our choices for the stellar IMF and the
cluster mass function. The methods of sampling these distribution
functions will be the topic of the next section.

\subsection{The stellar initial mass function}

For stars we will use the \citet{salpeter55} IMF:

\begin{equation} \label{eq:powerlaw}
\xi (m) = \frac{{\rm d} N}{{\rm d} m} = A \cdot m^{-\alpha},
\end{equation}

\noindent with $-\alpha = -2.35$. The reason for this choice is
computational simplicity. The steepening of the IGIMF as found by
\citet{kroupaweidner03, weidnerkroupa04} happens at relatively high
stellar masses, for which other IMFs \citep[e.g.][]{kroupa01,
chabrier03} have similar power-law indices. The differences  are
expected to be small between different IMFs. We will compare  the
Salpeter IMF to the \citet{kroupa01} IMF in Sect.
\ref{sec:samplingmethods}. The normalization constant ($A$) is
calculated from the total number or mass of stars. The minimum and
maximum stellar masses are taken to be 0.1 and 100 \msun, respectively.
Although there are indications that there is a fundamental  stellar
upper mass limit of $\sim$150 \msun\  \citep[][and references
therein]{weidnerkroupa04}, the upper stellar mass  limit has little
influence on our results.

\subsection{The cluster mass function}

For the star clusters we assume a power-law mass function, similar to
Eq.~\ref{eq:powerlaw}: 

\begin{equation}
\frac{{\rm d} N}{{\rm d} M} = B \cdot M^{-\beta}
\end{equation}

\noindent 
There exists a debate between different groups who try to obtain the cluster
initial mass function (CMF) in distant galaxies. In studies which try to
constrain the power-law slope of the CMF from the relation between the SFR of
a galaxy and the number of clusters in a galaxy (or, equivalently, the
luminosity of the brightest cluster in a galaxy), many groups find values of
$\beta = \sim 2.3-2.4$ \citep[e.g.][]{larsen02, whitmore03, weidner04,
gieles06a}. More direct measurements of the masses of the clusters, however, tend
to find values consistent with $\beta = 2.0$ \citep[e.g.][]{zhangfall99,
degrijs03, mccradygraham07, larsen08}. \citet{bastian08} notes that this
discrepancy can be alleviated by assuming that the clusters really follow a
Schechter-like mass distribution, which is a power-law at low masses, but
turns over at a typical mass into an exponential fall-off of the number of
clusters. The high mass of this turn-over (few $10^6$\msun) makes it hard to
infer directly from the masses. Their strong effect on the upper mass limit
for the clusters in a galaxy makes it detectable from a statistical point of
view, though. See below for a discussion on how Schechter-like CMFs might
influence the IGIMF effect.

Here we take pure power-laws with a slope of $\beta = 2.2$, for consistency
with the work of \citet{weidnerkroupa04}, and to have a case that is in
between  the values found by the two competing camps. In Sect.~\ref{sect:beta2} we discuss the specific case  $\beta = 2.0$, as well as a continuum of slopes in the range $\beta = 1.8 - 2.4$, in order to cover the whole range of slopes found observationally.

Although the range of cluster masses probed
is large, the observationally accessible extragalactic star clusters
have masses exceeding 1000 \msun, except for clusters in the Magellanic
Clouds. As minimum mass for star clusters we use a default value of 5 \msun, 
as did \citet{weidnerkroupa04}. As the value for a
physical lower mass limit for clusters, if any, is unknown, this mass
is taken,  because it is the lowest mass of groups of stars that is
observed to be  forming in the Taurus-Auriga region \citep{briceno02}.
This lower limit is far below
the range in which the power-law behavior is observed. It is an
extrapolation of more than two orders of magnitude. This extrapolation
is assumed in other IGIMF studies as well, and the best we can
currently do. The upper mass limit for star clusters is set to infinity. 

We vary both the  lower and the upper mass limits to investigate how
sensitive our results are to variations in these values. The minimum
cluster mass is expected to be important, and 5 \msun\ is far below
observational limits of any young star cluster that is outside the
solar neighbourhood. Observational indications for an upper cluster
mass limit are found in e.g. the Antennae \citep{zhangfall99} and M51
\citep{gieles06, haas08} and in general from the relation between the
brightest cluster in a galaxy and its star formation rate by
\citet{weidner04, bastian08}. These upper mass limits are found to be
around $10^{5.5 -  6.5}$ \msun. See Section~\ref{sec:sfrdependence} for
an investigation of star formation rate dependent IGIMFs.

\section{Sampling techniques} \label{sec:sampling}

In this section we discuss several ways to sample the distribution functions 
described in the previous section.

\subsection{Star formation scenarios and sampling of the IMF}

Ideally, one would like to connect sampling methods in numerical
experiments like the one conducted here in some way to the astrophysics
going on in the studied system. Here, this would mean that we construct
a method of sampling stellar masses in a cluster, which is based on a
scenario about how this cluster forms from its parent molecular cloud.
It is expected that the IMF found in star forming regions harbours a
wealth of information about the star formation process. A recent paper
by \citet{dib09} indeed describes several ways of building up an IMF
from star formation scenarios.

The problem with constructing sampling methods in this way, is that it
is not at all guaranteed that the mass function inside clusters follows 
the same functional form in all clusters. Besides, the mass function of
cloud cores is an equally uncertain factor. Likewise, the large
number of free parameters and inherent uncertainties of physical star
formation scenarios would inhibit us to draw any conclusions. The point
of this paper is to show the effects of different sampling methods,
given that the underlying IMF is the same. We chose, therefore, to use a
single underlying IMF, and construct sampling methods that do not
necessarily represent physical star formation scenarios.

\subsection{Analytic sampling} \label{sec:analytic}

The first method to sample a distribution function we discuss is
analytic in nature. We use the fact that the total mass of stars inside
a cluster (i.e. the cluster mass) is calculated from

\begin{equation} \label{eq:clustermass}
M_{\textrm{\tiny cl}} = \int_{m_{\textrm{\tiny min}}}^{m_{\textrm{\tiny max}}} m \cdot \xi(m) \, {\rm d} m
\end{equation}
where $m_{\textrm{\tiny min}} = 0.1$ \msun\ and $m_{\textrm{\tiny
max}} = \textrm{min}(100 \msun,\, M_{\textrm{\tiny cl}})$. Limiting the
mass of the most massive star present in the cluster ensures
that there are no stars more massive than their host cluster.

The normalization of the IMF ($A$ in Eq.~\ref{eq:powerlaw}) is defined by
relation~\ref{eq:clustermass}. Sampling the distribution function is
done by using 

\begin{equation} \label{eq:number}
N_i = \int_{m_{i}}^{m_{\textrm{\tiny max}}} \xi(m) \, {\rm d} m 
\end{equation}

\noindent
with $N_i = N_1$ = 1 for the most massive star (this star has the mass
$m_1$), 2 for the second most massive and so on. For any cluster mass 
the masses of all stars present in the sample are uniquely determined, 
see also \citet{weidnerkroupa04}.

\subsection{Random sampling}

In order to introduce stochastic effects, we will mainly sample mass 
functions randomly, as it ensures that random fluctuations are present
in the sample of masses. Whereas the analytic way of sampling will never
produce a 80 \msun\ star in a 100 \msun\ cluster, this will happen
(although rarely) when sampling randomly. There are nevertheless issues,
as described below.

A random number from a distribution function is drawn using a random 
number, uniformly distributed between 0 and 1 as many numerical 
packages can provide you with, and the normalized cumulative 
probability function, which is the normalized cumulative probability
density function, which (in this case) itself is an integral over the
mass function :

\begin{equation} \label{eq:cpdf}
\textrm{CPDF}(m) = \int_{m_{\tiny{min}}}^m \textrm{CMF} \, {\rm d} m
\end{equation}

\noindent normalized to \textrm{CPDF}($m_{\scriptsize{max}}$) $\equiv$ 1.
Inverting Eq. \ref{eq:cpdf} and inserting uniformly distributed
random  numbers provides the desired randomly sampled masses.

For power-law distribution functions, the inversion  can be done
analytically, such that the necessity for time consuming numerical 
integration or the use of look-up tables (constraining the flexibility
of our research) is prevented.

\subsubsection{The total mass of the cluster}
\label{sec:totmass}

When sampling stars one by one, the chances of them adding up to exactly
the cluster mass are marginal. Therefore, one has to make a choice about
which stars to include. One way is just sampling stellar masses until
you first go over the predetermined total cluster mass. Four choices
can be made:

\begin{enumerate}

\item Stop at that point. The cluster mass will always end up slightly
higher than the predetermined value. We will indicate this method by 
`stop after', as we always stop just after passing the cluster mass aimed 
for.

\item Remove the last star drawn. The cluster mass will now be
systematically lower than the masses drawn from the CMF, we will therefore 
abbreviate it by `stop before'.

\item Only remove the last drawn star if then the total mass is closer
to the desired value. The cluster masses are sometimes slightly
lower, sometimes slightly higher than the predetermined value.
This will be our default choice, indicated by `stop nearest'.

\item Like the previous option, but removing the star at 50\% probability, 
regardless of whether it would bring the cluster mass closer to the 
predetermined mass or not. This will be called `stop 50/50'.

\end{enumerate}

\subsubsection{Sorted sampling \`a la WK06} 
\label{sec:sortedsampling}

An alternative treatment was introduced and extensively tested by
WK06, `sorted sampling':  Draw a number of stars ($N = M_{\textrm{\tiny
cl}}/m_{\textrm{\tiny average}}$) in which $M_{\textrm{\tiny
cl}}$ is the cluster mass and $m_{\textrm{\tiny average}}$ is the average stellar mass in the IMF under consideration. Then draw that many stellar masses from the IMF. Repeat to do so if the total mass is not yet the desired
cluster mass, by drawing an additional $(M_{\textrm{\tiny cl}} - \sum_i
m_i)/m_{\textrm{\tiny average}}$ stars (where $\sum_i
m_i$ is the sum of the masses already drawn). When the cluster mass is first
surpassed, sort the masses ascendingly and remove the most massive star
if that brings the total stellar mass closer to the desired cluster
mass. Only the most massive star drawn can be removed. If the first sample of stars goes over the cluster mass by a large amount, still only one star can be removed, while the correction upwards in mass can be with any arbitrary number of stars.

\subsubsection{Sampling to a total number of stars}

Alternatively, one can draw once a predetermined number of stars for a
given cluster from the IMF. The number of stars that is drawn is, as in `sorted sampling', given by $N = M_{\textrm{\tiny
cl}}/m_{\textrm{\tiny average}}$. In this case some clusters may become much
more or much less massive than  the mass that was sampled from the
cluster mass function. We will indicate  this method with simply
`number'.

\begin{figure}
\centering
\includegraphics[width=\columnwidth]{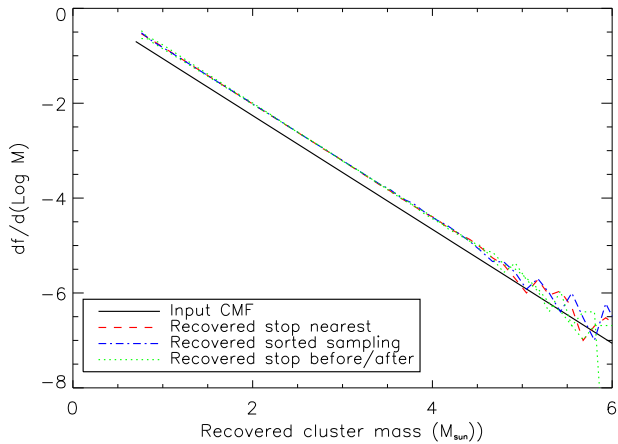}

	\caption{The fraction of clusters per unit $\log(M)$ as a function of 
	cluster mass. The input CMF is shown as solid straight black 
	line. The coloured, discontinuous lines are recovered CMFs after 
	populating clusters with stars with the indicated sampling methods.
	The input CMF is plotted offset, to more easily distinguish the 
	recovered CMFs.
            }
      \label{fig:recoveredcmf}
\end{figure}

\begin{figure}
\centering
\includegraphics[width=\columnwidth]{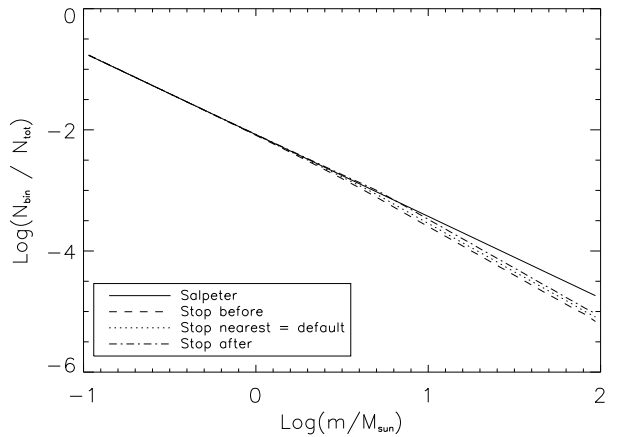}

	\caption{The IGIMF for randomly sampled stars in clusters until (1)
the next star would overshoot the cluster mass (dashed), (2) a mass
nearest to the cluster mass is reached (dotted), or (3) one star crosses the
cluster mass (dot-dashed). The solid line is the input Salpeter IMF. The
value on the vertical axis is the fraction of all the stars that are in
that particular mass bin.} \label{fig:igimf_rs} \end{figure}

\subsubsection{Limiting the stellar masses to the cluster mass}

By default we limit our maximum possible stellar mass to the mass of the
cluster (so that, e.g., a $M_{\textrm{\tiny cl}}$ = 10 \msun cluster can contain only stars at most as massive as $m_{\textrm{\tiny max}}$ = 10 \msun). Otherwise,
clusters of a predetermined mass may end up with a star that is more
massive than the cluster itself. However, we will also try without this
constraint, in which case we add `unlimited' to the name. Note that lowering the maximum possible stellar mass makes the probability for drawing lower mass stars (per unit mass) larger, as the integral of the probability density function of stellar masses should still be one.

\begin{figure*}
\centering
\includegraphics[width=\textwidth]{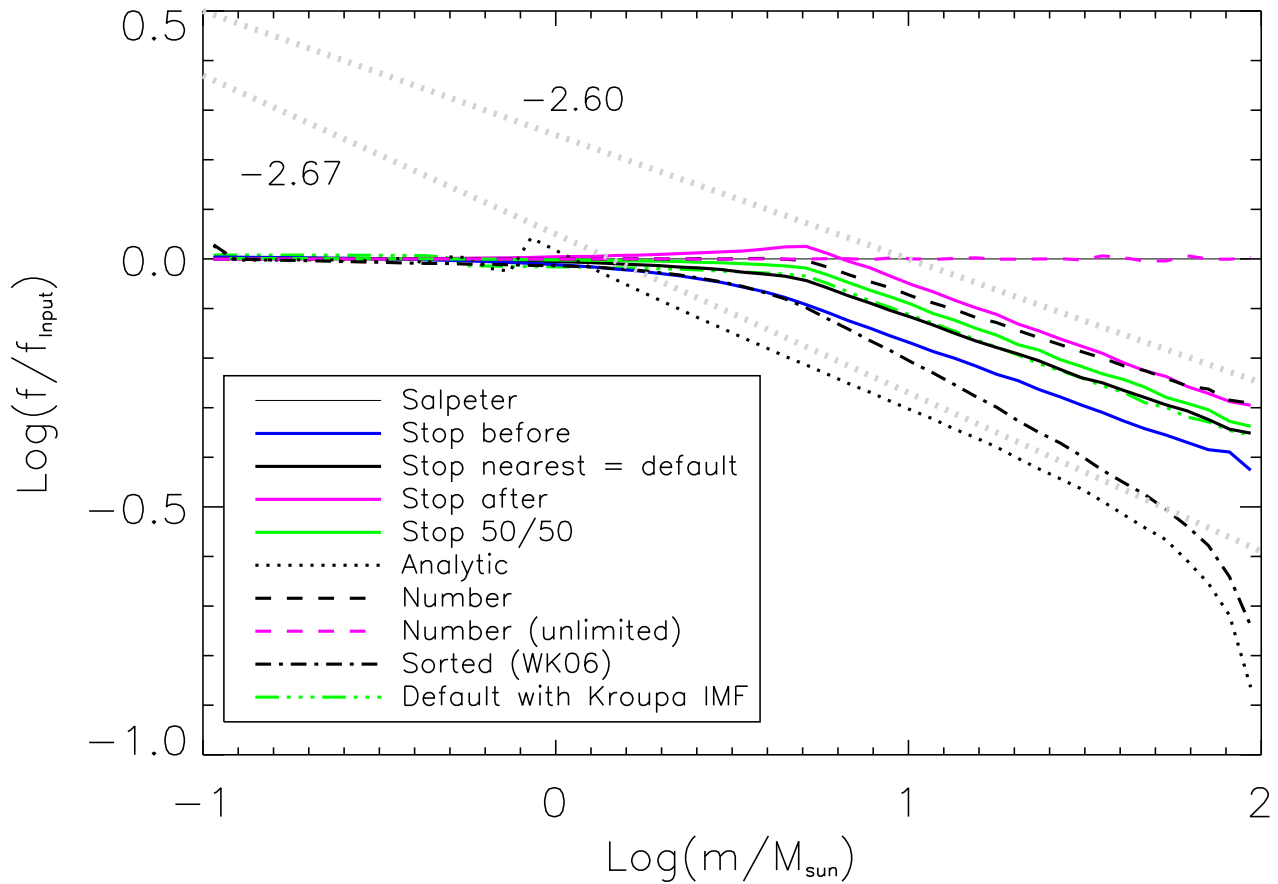}

	\caption{The same as Fig.~\ref{fig:igimf_rs}, but for every mass bin
divided by the expected value for the input IMF. The IGIMFs presented in 
Fig.~\ref{fig:igimf_rs} are represented by the solid lines, in various 
colours. The `stop 50/50' model is included as well. We also include the
`analytic sampling' case (dotted line), the sampling of a
specific \textit{number} of stars based on the expected mean mass,
limited  by the cluster mass (black dashed) and unlimited (magenta dashed,
going around the Salpeter line). The black dot-dashed line is the `sorted
sampling' method of WK06. The realisation for a \citet{kroupa01} IMF is
shown in the green dot-dot-dot-dashed line (almost on top of the black solid (default)
line). The light grey dotted lines with numbers are lines that would
have the indicated power-law index in the IGIMF.}

      \label{fig:ignorm_rs}
\end{figure*}

\subsection{The recovered cluster mass function}

One consistency test for the sampling methods is to see whether or not the mass
function of the clusters after populating them with stars from the IMF
recovers the input CMF. For some of the methods mentioned it is obvious
that the total mass will always be over- or underestimated (e.g.,
stopping the sampling always right after or right before you passed the
cluster mass, where the mass will  be over- or underestimated by on
average half an average stellar mass for  that IMF). For the high
cluster mass end these difference are negligible,  but that is not
necessarily clear for very low cluster masses, where the  recovered CMF
could be steeper or shallower than the input CMF.

In Fig.~\ref{fig:recoveredcmf} we compare the input CMF (solid black
line,  shifted by an arbitrary vertical offset), to several recovered
CMFs after populating the clusters with stars. The default sampling
method is shown in red  (dashed), and the preferred method of WK06,
sorted sampling, is shown in  dash-dotted blue. The two models for which
discrepancy is expected are shown  in the dotted green lines. The
expected under- or overestimate of the total  mass is $\sim0.3$ \msun,
which is more than an order of magnitude less than the very lowest
cluster mass. It turns out that even for these models the  discrepancy
is marginal. Therefore we cannot rule out one or another sampling 
methods based on the recovered CMF.

\begin{figure*}
\centering
\includegraphics[width=\textwidth]{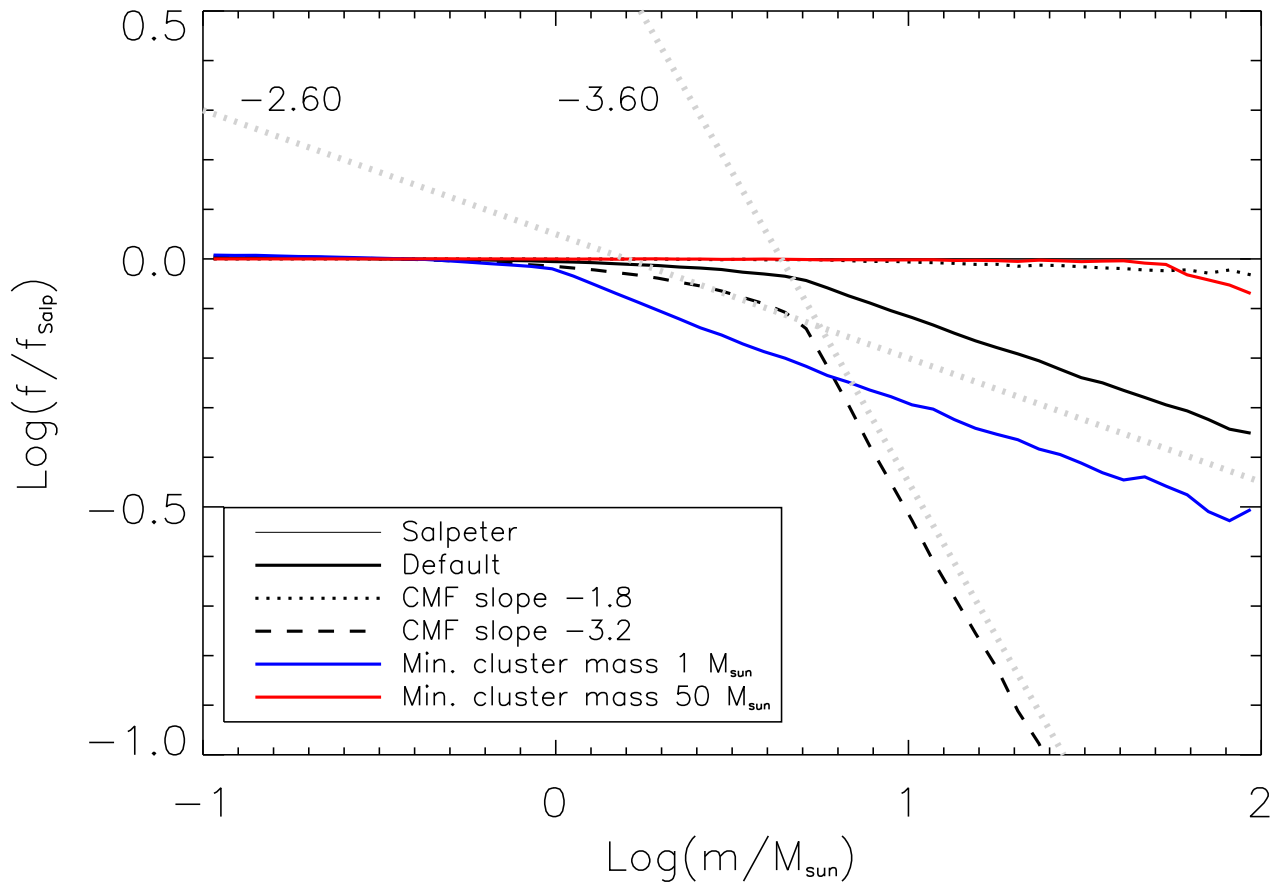}

	\caption{The same as Fig.~\ref{fig:ignorm_rs}, but now for variations
of the cluster mass function. We show our default (minimum mass 5 \msun,
power-law index -2.2) model and 4 other models: slopes varied to -1.8
(dotted) and -3.2 (dashed) and the minimum cluster mass set to
1 (blue solid) and 50 \msun (red solid). The light grey dotted lines give an indication of the slope of the lines, when plotted as an IGIMF, with power-law indices as indicated.
            }
      \label{fig:ignorm_rscmf}
\end{figure*}


\section{Integrated Galactic Initial Mass Functions} \label{sec:results}

We draw samples of $10^7$ clusters from a cluster mass function with ${\rm d}
N/ {\rm d} M \propto M^{-2.2}$. We tested several sample sizes and found
$10^7$ to be both computationally feasible and showing only tiny
statistical fluctuations (using, e.g., $10^6$ clusters results in IGIMF
scatter nearly as big as the difference between some models we test). We
construct the IGIMF by sampling the stars in the clusters in different
ways, as described in the previous section, and sum up all stars from
the individual clusters. In Fig.~\ref{fig:igimf_rs} we show three IGIMFs
from random sampling, together  with the Salpeter IMF.

\subsection{Sampling methods}
\label{sec:samplingmethods}

Figure~\ref{fig:igimf_rs} clearly indicates that the IGIMF steepens for
high stellar masses, due to the lack of high-mass stars in low-mass
clusters. Also, the impact of using either of the three methods is
comprehensible: stopping the sampling one star before the cluster mass
is filled up biases most against high stellar masses (as the chance of
going over the cluster mass is higher for a higher-mass star) and going
slightly over the cluster mass biases least against high mass stars.
Because the differences are small, from now on we plot the fraction of
all stars in a mass bin, divided by the fraction predicted from
the input stellar IMF (i.e., \citet{salpeter55}). The same data as in
Fig.~\ref{fig:igimf_rs} are used for Fig.~\ref{fig:ignorm_rs}, where the
differences become clearer.

In Fig.~\ref{fig:ignorm_rs} we also compare the analytic method of
sampling, as explained in Section~\ref{sec:analytic}, to the random
sampling methods. Both have the stellar masses limited to be at most the
cluster mass, but in the random sampling technique sometimes a
relatively high-mass star does occur in a low-mass cluster. This is not
the case using the analytic sampling, resulting in the sharp downturn 
at masses close to the upper limit. As noted by WK06, the `sorted 
sampling' method resembles the shape of analytic sampling, although 
less severe. The relation is even steeper (approaching an IGIMF 
power-law index of -3) for 
$m>M_{\textrm{\tiny cl, min}}$.

Sampling a number of stars equal to the cluster mass divided by the
average stellar mass for the IMF under consideration is also shown in
Fig.~\ref{fig:ignorm_rs}. If the average mass is calculated with the
upper mass limit in a cluster limited to the cluster mass, then the
method gives results rather similar to the default method. When the
average mass is always calculated for a well sampled IMF between 0.1 and
100 \msun\ then the resultant IGIMF is indistinguishable from the input
IMF. Note that the cluster mass function is still intact.

Using a \citet{kroupa01} IMF results in the green solid line in 
Fig.~\ref{fig:ignorm_rs}. The bend again is found at roughly the 
same stellar mass as for the Salpeter IMF. Deviations from this at lower mass 
are stronger, though, as the mean mass of a star in the Kroupa IMF is bigger
than in a Salpeter IMF. Changing the upper stellar mass limit does not 
influence any of the results other than that the lines extend to higher 
stellar masses.

Comparing the calculations to the light grey dotted lines in
Fig.~\ref{fig:ignorm_rs} shows that all random sampling techniques give
high-mass-end power-law indices of the IGIMF very close to -2.60,
whereas the analytic sampling technique is slightly steeper, $-2.67$, and
turning completely down close to the physical upper mass limit for stars
(i.e. the mass of the cluster needs to become extremely high in order to
sample a star with a mass very close to the upper mass limit).

\subsection{The cluster mass function}

In all randomly sampled realizations, the steepening becomes very
prominent at $m = 10^{0.7}$\msun\ $ = 5$ \msun, the lower mass limit
for clusters. The analytically sampled case becomes steeper at lower
masses, as there the most massive stars in these low-mass clusters are
well below the cluster mass. We investigate how the steepening
depends on the imposed lower cluster mass limit and the steepness of the
cluster mass function.

In Fig.~\ref{fig:ignorm_rscmf} we show the IGIMFs, as obtained with our
default random sampling model, for three different lower cluster mass
limits (1, 5 and 50 \msun, for a CMF slope of -2.2) and three different
cluster mass function slopes (-1.8, -2.2 and -3.2 , for a lower
cluster mass limit of 5 \msun). The flattest CMF and highest minimum
cluster mass use samples of $10^6$ clusters instead of $10^7$ clusters. The
higher mass in clusters makes  the IGIMF less sensitive to errors from
sampling statistics in stars.

It can be clearly seen that the deviations of the IGIMF from a standard
Salpeter IMF start at the minimum cluster mass. Results therefore are
sensitively dependent on the cluster mass functions at very low cluster
masses. The steepness of the IGIMF depends on the power-law slope of the
cluster mass function. Changing the cluster mass function power-law
index from -2.2 into -3.2 (-4.2, not shown in the figure) makes the IGIMF slope steepen from -2.6
to -3.6 (-4.8). The steepening still occurs at the minimum allowed
cluster mass. A flatter CMF slope of -1.8 results in a much shallower
IGIMF compared to our standard case, with little deviations from the
input \citet{salpeter55} IMF. We can also conclude 
that, as long as the {\sl lower} mass limit of the CMF is larger than 
the {\sl upper} mass limit of the IMF, IGIMF effects are negligible. We 
would like to emphasize that, although we used a lower mass limit of the 
CMF of 5 \msun\ (i.e. considerably lower than the upper mass limit of the IMF), this 
value as well as the shape of the CMF at masses below a few hundreds 
solar masses is very uncertain due to a lack of observational data, even 
in the Milky Way.

\subsection{The $\beta=2$ CMF}
\label{sect:beta2}
\begin{figure}
\centering
\includegraphics[width=\columnwidth]{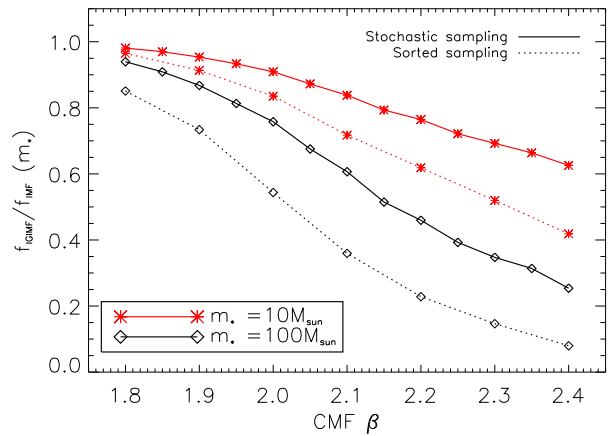}

\caption{The deviation at 10 and 100\msun (stars and diamonds, respectively) of the value of the IGIMF as compared to the IMF as a function of the CMF slope $\beta$ in the region around $\beta=2$. For the `sorted sampling' method (dotted lines) we performed the Monte Carlo simulations at intervals of $\Delta \beta = 0.1$ from 1.8 to 2.4 (including the entire range of observationally determined values) using one million clusters. The default, fully stochastic sampling (solid lines) simulations are performed using ten million clusters with $\beta$ varying steps of 0.05.
}
\label{fig:around2}
\end{figure}

In \citet{elmegreen06} there is the claim that in the case where the CMF is described by a power-law of $\beta=2$, IGIMF effects vanish, making this a singular case in between our $\beta=1.8$ and $\beta=2.2$ cases. In order to validate this result, we run simulations with values for $\beta$ close to, and including, 2. In order to address the behaviour of the deviation of the IGIMF from the IMF, we plot the deviation of the IGIMF from the underlying IMF at two different stellar masses as a function of $\beta$ in Fig.~\ref{fig:around2}. We use a minimum mass of 5\msun\ for the cluster CMF, no upper mass limit and plot the results for $m_*$ = 10 and 100 \msun\ in bins of width $\Delta \log(m_*) = 0.2$. We perform the exercise for our default sampling method at 13 different values of $\beta$ (i.e., with $\Delta \beta = 0.05$), and for 7 different values with the `sorted sampling' technique from WK06 (i.e., with $\Delta \beta = 0.1$). 

We find the results for $\beta=2$ to be non-singular and to follow the expected behaviour from its surrounding points. The vanishing effects found by \citet{elmegreen06} are not reproduced in our simulations. In the Monte Carlo simulations described in \citet{elmegreen06} a small difference was already visible. In the intuitive analytic section it is explained why there should not be a difference. This conclusion is based on the crucial statement that `the probability of forming a star of a particular mass is independent of cluster mass'. This is only true for stars in clusters with masses higher than the upper mass limit for stars. For clusters with lower total masses, the situation is more complex: stars with masses higher than the total cluster mass get assigned zero probability (unless one does not impose a limit to the stellar mass equal to the cluster mass), while stars with lower masses get higher probabilities to fulfil the IMF normalization. For any value of $\beta$ there is some number of clusters which will lack high mass stars, which makes $\beta=2$ a normal case without singular features. The claim by \citet{elmegreen06} is correct only if the lower 
limit of the CMF is higher than the maximum stellar mass, in agreement with our own findings.

We learn from Fig.~\ref{fig:around2} that choosing a value for the power-law index of the CMF of -2.2 instead of -2.0 gives a larger effect, as does the choice of sampling method made by WK06, compared to our default method. The observational support for $\beta=2.4$ justifies the use of $\beta=2.2$ in the rest of this work.

\subsection{Star Formation Rate dependent upper cluster mass limit}
\label{sec:sfrdependence}
On purely statistical grounds a relation between the star formation rate 
(SFR) and the mass of the most massive cluster in a galaxy is expected, 
as long as the timescale to form a complete population of clusters is 
fixed \citep[see][they find a timescale of 10 Myr]{maschbergerkroupa07}. This relation is expected to be linear.
However, \citet{weidner04} have found a relation between the SFR of a 
galaxy and the mass of its most massive cluster that is slightly shallower 
than linear:
\begin{equation} \label{eq:sfrdep}
\log(M_{\scriptsize{cl, max}}) = 0.75 \cdot \log(\textrm{SFR}) + 4.83
\end{equation}
In this section we will show Monte Carlo simulations with upper cluster 
mass limits that correspond to SFRs of $10^{-5}$ to $10^{4}$ 
\msun\ yr$^{-1}$ in steps of half a dex in SFR.

As galaxies with a very low SFR in general also have very low masses 
(dwarf galaxies), we can expect to have more statistical (shot) noise in
low SFR samplings. In order to get a handle on the median IGIMF and the 
68\% ($\sim 1 \sigma$) confidence intervals we assume that the galaxies 
have formed stars for 10 Gyr, together with the SFR this gives a total 
stellar mass. The CMF (with a lower mass cut-off of 5\msun\ and a
power-law  index of -2.2) then sets the number of clusters drawn. For
the very low SFR  runs, there are not so many clusters to be drawn
($10^{-5}$ \msun\ yr$^{-1}  \cdot 10$ Gyr $= 10^5$ \msun\ total stellar
mass) we run 500 realizations of  the lowest SFRs, gradually reducing
this number as the 68\% confidence  intervals are very narrow, already
for relatively low SFRs. The corresponding upper cluster mass limits
range from 10$^{1.08}$ = 12 \msun\ to 10$^{7.83}$ \msun, 
so extending from  extremely (maybe even unphysically) low star
formation rates and corresponding upper cluster mass limits to
extremely high SFR limits. Both limits are far  beyond the range in
which the relation between SFR and maximum cluster mass  has been
observed. We sample the IMF using the method which samples up to a
total mass  and removes the last drawn star if that brings the total
mass of stars  closer to the predetermined cluster mass described
before (i.e., ``stop nearest'').

\begin{figure}
\centering
\includegraphics[width=\columnwidth]{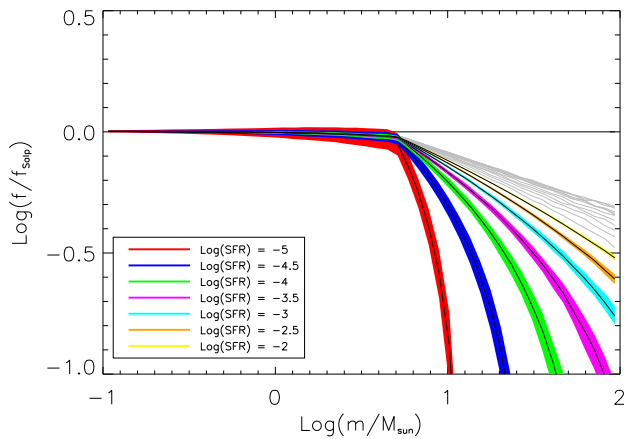}

	\caption{SFR dependent IGIMFs, in which the SFR sets an upper cluster mass 
	limit, given by Eq.~\ref{eq:sfrdep}. We ran the lowest SFR models sufficiently 
	long to get converged confidence intervals which are shown by the coloured 
	regions around the solid line medians. For the higher SFR simulations, the 
	results are very close together, and the confidence intervals extremely narrow. 
	Therefore, we only plot the result of 1 simulation . The order is such that 
	the higher the star formation rate (and hence the upper cluster mass limit), 
	the shallower the IGIMF. Note that the highest SFR run is $10^4$ \msun\ yr$^{-1}$.
            }
      \label{fig:sfrdep}
\end{figure}

In Fig.~\ref{fig:sfrdep} we show the IGIMFs for the 19 different SFRs
(solid lines). For the simulations with a series of runs we show medians
(in black) and 68\% confidence intervals in colour. It appears that for
a given CMF and sampling method the statistical variation around the
median IGIMF is very  small. Also, for lower SFRs the high mass end of
the IGIMF is steeper, due to  the lower upper cluster mass limit. With a
lower upper cluster mass limit  relatively more clusters form with a low
mass. As the upper cluster mass limit  increases, the variation in the
IGIMF becomes smaller. This  indicates that our simulations, without an
upper limit, are good representatives for high SFR objects (galaxies),
whereas for galaxies with a low SFR the IGIMFs are steeper. So, for galaxies with a low SFR,  
\textit{the effect will in reality be stronger than we indicate}.

In \citet{bastian08} it is claimed that in order to reproduce the
relation between SFR and the maximum cluster luminosity, it is preferred
to have a Schechter-like CMF (i.e. a power-law with an exponential
cut-off above some mass) instead of a pure power-law. The typical mass
at which the CMF turns down exponential is a few times $10^6$\msun.
As this mass is much higher than the highest stellar mass, the
precise shape of the cut-off is not expected to be important. An
exponential turn-down at that mass has a similar effect on the IGIMF to
truncating the CMF at that mass. For the lower limit to the cut-off
mass found by \citet{bastian08} the SFR corresponding to their cut-off
mass, according to Eq.~\ref{eq:sfrdep}, would be $10^{1.6}$ \msun\
yr$^{-1}$. In Fig.~\ref{fig:sfrdep} it can be seen that such IGIMFs are
hardly distinguishable from CMFs without upper cluster mass limits.

\subsection{Constructing IGIMFs from clustered and non-clustered star formation}
The results described above are only valid if all stars are born in clusters. The fraction of stars formed in clusters is a strongly debated quantity nowadays. Different authors constrain the fraction of stars formed in clusters in different, not necessarily comparable, ways. The main hindrance here is the definition of `a cluster'. Young clusters often get disrupted (sometimes called `infant mortality') on time scales of about $10^7$ yrs \citep{tutukov78, kroupa01a, lamers05}. These young clusters might, or might not, have a stellar mass distribution similar to clusters which survive their childhood. Also, stars may form without ever being part of a ``cluster''. Numbers for the estimate of the fraction of stars born in clusters vary from $\sim 5-10\%$ \citep[][ and references therein]{millerscalo78, bastian08} up
to 40\% or higher as found in the comparison of cluster mass production for a
particular CMF power-law index by \citet{piskunov06}. Different authors use different definitions of what a cluster/association is and find very different values for the fraction of stars that is a born in a cluster-like environment \citep[see e.g. also ][]{carpenter00, ladalada03, porras03, megeath05, piskunov08}.

The `real IGIMF' (i.e. the true distribution of stellar masses at birth for a whole galaxy) can be straightforwardly estimated from the IGIMF from clustered star formation (i.e. the results given above), and the IMF from stars born in isolation. If we denote the distribution of initial masses in the field as IMF$_{F}$, the IGIMF from clustered star formation (i.e. the results obtained above) as IGIMF$_C$ and the total IGIMF (the pdf of initial masses of all stars in a galaxy) as IGIMF$_T$, we can simply write at any given stellar mass:
\begin{equation} \label{eq:igimf_tot}
{\rm IGIMF}_T (m_*) = f \cdot {\rm IGIMF}_C(m_*) + (1-f)\cdot {\rm IMF}_F (m_*)
\end{equation}
where $f$ is the fraction of the stellar mass that is born in clusters, assuming that this fraction $f$ is independent of stellar mass and that the mass distributions in the right-hand side of the equation refer to distributions which are both well sampled. In practice, this means that the total IGIMFs will end up in between the IGIMFs described above and the underlying IMF, weighed by the fraction of clustered star formation (so lines in Fig.~\ref{fig:ignorm_rs} and \ref{fig:ignorm_rscmf} will end up in between the horizontal line and the shown IGIMFs).

Note that, if the second term in the right-hand side of Eq.~\ref{eq:igimf_tot} is large, that IGIMF effects may well become negligible, or at least far less significant than indicated in the rest of this paper.


\section{The number of O-stars in the Milky Way} \label{sec:Ostars}

One way to judge between the several IGIMFs (or judging on the
importance of the IGIMF effect) would be high mass star counts
by upcoming surveys like GAIA \citep[e.g.][]{perryman01}.

In order to estimate how many O-stars will be observed by GAIA, we
will here do the following exercise, in which we keep things as simple
as possible. We assume that the IGIMFs described earlier are perfectly
sampled (i.e. there are no sampling issues besides the ones that make
up the IGIMFs), that the SFR of the Milky Way has been constant for the
last 10 Myr, which we will assume to be the lifetime of O-stars.
Furthermore, we assume that the fraction of all O-stars in the Milky
Way, observed by GAIA, is the same as the fraction of all stars
together (i.e. $\sim$10\%). This last number is very uncertain. O-stars
are very bright and would therefore be visible to larger distances (the
GAIA survey will be magnitude limited). If, however, all O-stars form
in the disc, the extinction towards them will be typically higher than
for stars above and below the disk. A fraction of the O-stars maybe
runaway stars, launched by multiple body interactions in young star
clusters, which can bring them from the disk into less dusty regions
(O-stars formed in isolation will typically not get far out of the
disk, as with a random velocity of a few times 10 km/s, they will not
get much further than a few parsecs away from the disk plane they were
formed in). The observed number of O-stars is then given by

\begin{equation} \label{eq:nostars}
N_O = A_\textrm{\scriptsize IGIMF} \, \Big(\frac{\textrm{SFR}}{1 \textrm{\msun/yr}}\Big)\, \Big(\frac{\Delta t}{10 \textrm{Myr}}\Big) \, \Big(\frac{f_\textrm{obs}}{0.1}\Big) 
\end{equation}

\noindent in which SFR is the SFR of the Milky Way, averaged over
$\Delta t$, which is the lifetime of O-stars and $f_\textrm{obs}$ is
the fraction of O-stars in the Milky Way that will be observed.
$A_\textrm{\scriptsize IGIMF}$ is the number of O-stars under the given
assumptions, calculated by dividing the total mass formed by the
average stellar mass of the IGIMF, multiplied with the fraction of all
stars that are more massive than 17 \msun, in which all the IGIMF
information is absorbed. In Table~\ref{tab:a_igimf} we give the factor
$A_\textrm{\scriptsize IGIMF}$ for the Salpeter IMF and all our IGIMFs.
We rounded the numbers off to multiples of ten.

\begin{table}
\caption{The factor $A_\textrm{\scriptsize IGIMF}$ from Eq.~\ref{eq:nostars} for Salpeter and all our IGIMFs, with underlying Salpeter IMF in the second column and a Kroupa IMF (for selected sampling methods) in the third column. These numbers are the number of O-stars found by GAIA, if the assumptions we explain in the text are correct, if the given IGIMF is the true IMF integrated over the galaxy. The numbers are rounded of to multiples of ten. The last column gives the ratio between the second and third column.}             
\label{tab:a_igimf}      
\centering                          
\begin{tabular}{l c c c}        
\hline                
\hline
IGIMF & $A_\textrm{\scriptsize IGIMF, Salpeter}$ & $A_\textrm{\scriptsize IGIMF, Kroupa}$ & ratio \\
\hline
Underlying IMF & 2610 & 4090 & 1.9 \\      
\hline
Stop nearest & 1650 & 2670 & 1.62\\
Stop before & 1490 & & \\
Stop after & 1830 & & \\
Stop 50/50 & 1710 & & \\
Analytic & 1050 & & \\
Number & 1810 & & \\
Number unlimited & 2610 & & \\
Sorted sampling & 1200 & 2050 & 1.71 \\
\hline
CMF slope -1.8 & 2530 & & \\
CMF slope -3.2 & 280 & 450 & 1.61 \\
CMF slope -4.2 & 60 & & \\
Min. cluster mass 1\msun & 1210 & & \\
Min. cluster mass 50\msun & 2570 & & \\
\hline                                   
\end{tabular}
\end{table}

It is now well established that the real IMF in star forming
regions is not Salpeter-like, but bends over towards lower masses
\citep[e.g.][]{kroupa01, chabrier03}. The difference here mainly lies
in the number of very low mass stars, for which our IGIMFs are all
indistinguishable from the underlying IMF. The fraction of O-stars in
IGIMFs with other underlying IMFs will be different though, as the
fraction of very low mass stars is lower than in Salpeter, making the
fraction of high mass stars higher. For example, the numbers in a
Kroupa or Chabrier IMF will be about 1.6 times higher (the exact values
of the ratio depends on the sampling methods and cluster mass
functions, but do not vary much). To illustrate this, we ran a
selection of our sampling methods also for an underlying
\citet{kroupa01} IMF, as displayed in the third column of
Table~\ref{tab:a_igimf}. The last column gives the ratio between the
results for an underlying Kroupa IMF and an underlying Salpeter IMF.
From the ratios (for rather `extreme' sampling methods) it can be seen
that they do not vary much from one sampling method to the other.

From the numbers in Table~\ref{tab:a_igimf} it is clear that in
principle several IGIMFs may be ruled out by the GAIA survey. The
difficulty in judging between several IMFs will be in the other numbers
quoted in Eq.~\ref{eq:nostars}. Some of the extreme IGIMFs can most
probably be ruled out with less exact knowledge of the other important
parameters. We want to stress here that the given numbers are only the
number of O-stars observed by GAIA, if the underlying IMF is Salpeter
and if the cluster mass functions assumed are the true mass
distributions of clumps of forming stars (as here they are heavily
extrapolated from the observed mass ranges of young clusters).

\subsection{Clusters consisting of one (O-) star}

\begin{figure}
\centering
\includegraphics[width=\columnwidth]{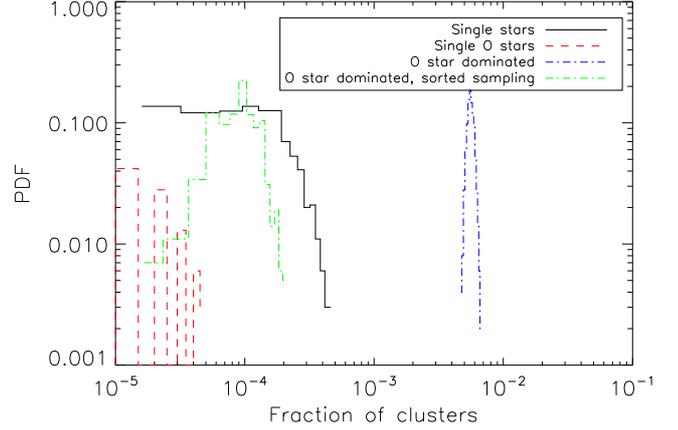}

	\caption{Distribution functions of clusters consisting of single
	stars (black solid line), single O-stars (red dashed line), and
	clusters dominated by O-stars (i.e. consisting of an O-star that has
	more than half of the  cluster mass, blue dash-dotted line) in the
	default sampling method, in a  cluster population with a power-law
	CMF with index -2.2 and a lower cluster  mass limit of 5\msun. In the
	sorted sampling method of WK06, the first two fractions are zero for
	all cluster samples. The O-star dominated fraction is 10 times lower
	than in the default method, as shown by the green dash-dotted  line.
	}

      \label{fig:singlestars}
\end{figure}

Using our sampling methods we might form clusters that consist of only
one  star. The question whether this is important or not was raised by
\citet{dewit04,dewit05}. We track here a) the fraction of clusters that
consist of a single star, b) the fraction of clusters that consist of a
single O-star ($m>17$\msun, see \citet{dewit05}) and c) the fraction of
clusters for which the most massive  star is an O-star, which contains
more than half the total cluster mass  (we call these ``O-star
dominated'' clusters). The results are shown in 
Fig.~\ref{fig:singlestars}. We plot probability distribution
functions (PDFs) for the fraction of clusters that have the indicated
properties in a cluster population. We ran ten thousand realisations of
cluster populations and counted, for example, how many clusters were
actually single O-stars, and divide that number by the total number of
clusters. The distribution of these fractions is what is plotted. So,
the peak of blue dot-dashed line shows that all ten thousand cluster
populations have about 0.5-0.6\% of their clusters being O-star
dominated. PDFs that do not add up to 1, like the fraction of clusters
that consist of exactly 1 O-star, indicate that the rest of the cluster
populations had zero single O-stars in them.

For the default sampling method, there are a few, but very little,
clusters that  consist of only one star, only 1 in $\sim10^4$. The
number of clusters in which  this one single star is an O-star is again
an order of magnitude smaller, with a median of $5.6\e{-3}$. This 
factor ten in between is less than the ratio of the number of O-stars
to the number  of all other stars, because it is more likely that you
are close to the mass of the  cluster if the star is an O-star, than
when it is less massive. The fraction of  clusters that is dominated by
an O-star (i.e. containing an O-star with at least  half the cluster
mass), shows a sharply peaked distribution function around 0.56\%. 

In the sorted sampling method of WK06, the fraction of clusters
containing a single  (O-) star is zero by construction: the first
number of clusters to be drawn is  calculated by dividing the total
cluster mass aimed for, divided by the mean stellar  mass in that
cluster, according to the appropriate IMF. This mean mass is more than 
an order of magnitude smaller than the assumed lower cluster mass
limit. Therefore,  of the order of ten stars or more are always drawn.
If the cluster mass is exceeded  already with the first drawing of
stars (for instance, if there is a really massive O-star drawn, that on
its own has as much mass as the rest of the stars or more), then at
most one star is removed, resulting in a cluster with at least on the
order of ten stars. The number of O-star dominated clusters is
therefore also much lower: the chance of having an O-star with half the
mass of the cluster or more, while not  going far over the cluster mass
(far enough to let that O-star be removed) is small.  10 stars will
mostly have an average mass that is close to the average mass of stars 
according to the IMF. The one dominating O-star then is several solar
masses too  massive, making it very likely to be removed. The median
fraction of O-star dominated clusters is $9\e{-5}$.

\subsection{The number of single O-stars}

In their paper, \citet{dewit05} specifically look at the fraction of
all O-stars that are single, i.e. not part of a detected cluster. They
claim very low mass clusters can be detected, such that these are
really O-stars without a surrounding cluster. Nevertheless they are
only sensitive to very low mass clusters if these clusters are very
concentrated (i.e. small). Clusters of very low mass are very easily
disrupted, and extrapolating the results of \citet{lamersgieles06} to
lower mass clusters (by about an order of magnitude), the typical
dissolution time of clusters is given by $t_d = 1.7\cdot
(M/10^4$\msun$)^{0.67}$ Gyr, resulting in O-star lifetimes for
$\sim10$\msun\ clusters. Therefore, it is very likely that if O-stars
live in very low mass clusters, the clusters are in the process of
being completely dissolved at the time of observation of the cluster.
If not completely disrupted yet, the cluster will have dispersed
already significantly, making it harder to detect the underlying
cluster than assumed by \citet{dewit05}.

We use this argument to claim that also our ``O-star dominated''
clusters would probably be observed as being single O-stars. Together
with the analysis of the previous section, we can now investigate what
fraction of all O-stars would be observed to live outside star clusters
(without taking runaway OB stars into account). For the default
sampling mechanism 11\% of the O-stars would be observed to live
outside clusters (if all O-star dominated clusters are detected as
single O-stars). For the sorted sampling method this is 0.24\%. The
difference of course is mainly caused by the different fraction of
O-star dominated clusters.

\citet{dewit05} found that 4$\pm$2\% of the O-stars in the Galaxy
cannot be traced back to a formation in a cluster or OB-association.
Although this number is smaller than what we find, taking into
consideration that we did include very low mass (and probably)
dispersed clusters it is legitimate to correct our result down by a
factor of a few, bringing the results in nice agreement. Increasing the
number of single O-stars in the ``sorted sampling'' method is much
harder to justify, so we conclude that that method significantly
under produces single O-stars, by a factor of 10-20.


\section{Galaxy evolution models}
\label{sec:galaxymodels}

The {\sc galev} models \citep{bicker04,GALEV09} are evolutionary
synthesis models for galaxies and star clusters. Essentially,
evolutionary synthesis models take a set of isochrones, assign a
suitable stellar spectrum to each isochrone entry, weigh each entry
according to a stellar mass function and a star formation history (SFH),
and sum up all contributions for a given isochrone age. {\sc
galev}s ``chemically consistent'' modeling follows the steady chemical
enrichment of the interstellar medium caused by stellar winds and
supernovae, and forms stars at the metallicity available in the gas
phase at this time. Nebular emission is taken into account for actively
star-forming galaxies.

We used models with the following input physics
\begin{itemize}
\item isochrones: from the Padova group (\citet{bertelli94} with
subsequent updates concerning the TP-AGB phase)
\item spectral library: BaSeL 2.2 \citep{lejeune97,lejeune98}
\item SFH as a function of Hubble type: following \citet{sandage86},
with parameters adjusted to reproduce simultaneously a range of
observations for galaxies of different Hubble types (for details see
\citet{GALEV09})
\begin{itemize} 
\item an Sd galaxy is modeled with a constant SFR
\item an E galaxy with an exponentially declining SFR with a 1/e decline
time of 1 Gyr
\item Sa-Sc galaxies are modeled with SFRs depending on the available
gas mass at a given time (similar to the Kennicutt-Schmidt law, see
\citet{kennicuttschmidt98}), resulting in approximately exponentially
declining SFR with 1/e decline times of 3.5 Gyr (Sa galaxy), 6 Gyr (Sb
galaxy) and 10.5 Gyr (Sc galaxy)
\item the gas mass-dependence of the Sa-Sc galaxies' SFR results in
slight changes between models with \citet{salpeter55} IMF and the
various IGIMFs, with the IGIMF models having slightly lower SFRs by
up to 5\% for our standard IGIMF model (``stop nearest'') and up to 10\%
for extreme cases
\end{itemize}
\item stellar yields: explosive nucleosynthesis yields are taken from
\citet{woosleyweaver95} for high-mass stars (M $>$ 10 \msun) and
from \citet{hoekgroenewegen97} for stars with lower masses. In addition,
SN Ia yields from \citet{nomoto97} are included (only total metallicity
is traced, not individual elements)
\item stellar MF: we use the various IGIMFs determined in this work
\end{itemize}

Underlying assumptions for this approach include
\begin{itemize}
\item the IGIMF does not change with time or SFR (taking into account
the SFR-dependent effects discussed in Sect. \ref{sec:sfrdependence}
would only strengthen the deviations, so our results are lower limits
for the impact of the IGIMF effect)
\item the IGIMF does not change with metallicity (no such dependence is
known or expected for Population I or Population II stars and star
clusters)
\item no infall or outflow of material is used (but also not needed
to reproduce a range of galaxy properties correctly, see
\citet{GALEV09}), likewise we neglect galaxy interactions
\item we assume instantaneous mixing and cooling of ejected material 
with the entire available gas reservoir (however, the SFH parameters
are adjusted to reproduce available gas metallicities as a function of
the galaxies' Hubble type at the present day)
\item we aim at modeling L$^*$ galaxies of the respective Hubble type,
hence neglect any magnitude-metallicity relation
\end{itemize}

For more details see \citet{GALEV09}.

\subsection{Integrated photometry of galaxies}

\begin{figure*}
\centering
\begin{tabular}{ccc}
\includegraphics[angle=270,width=0.45\linewidth]{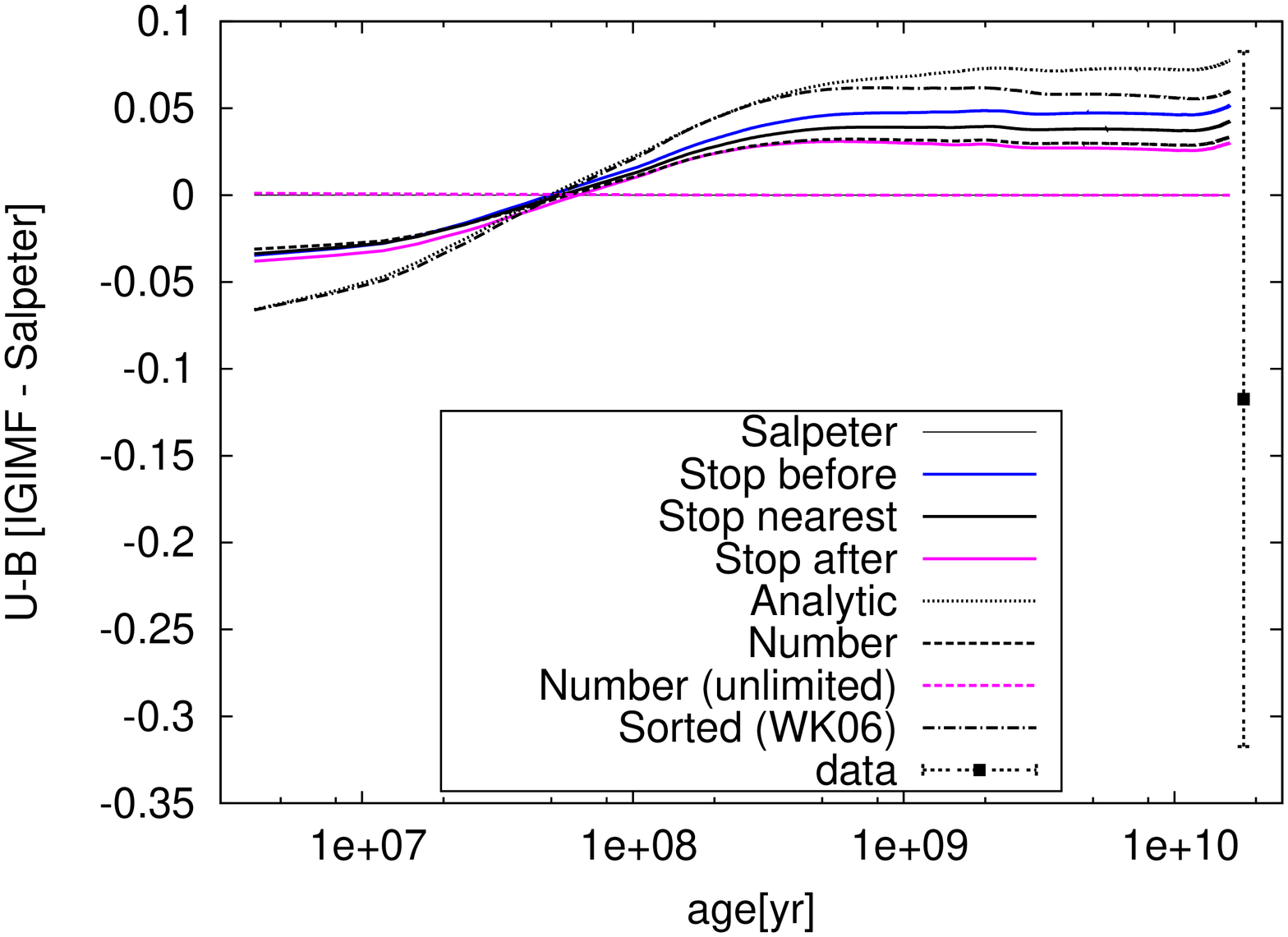} &&
\includegraphics[angle=270,width=0.45\linewidth]{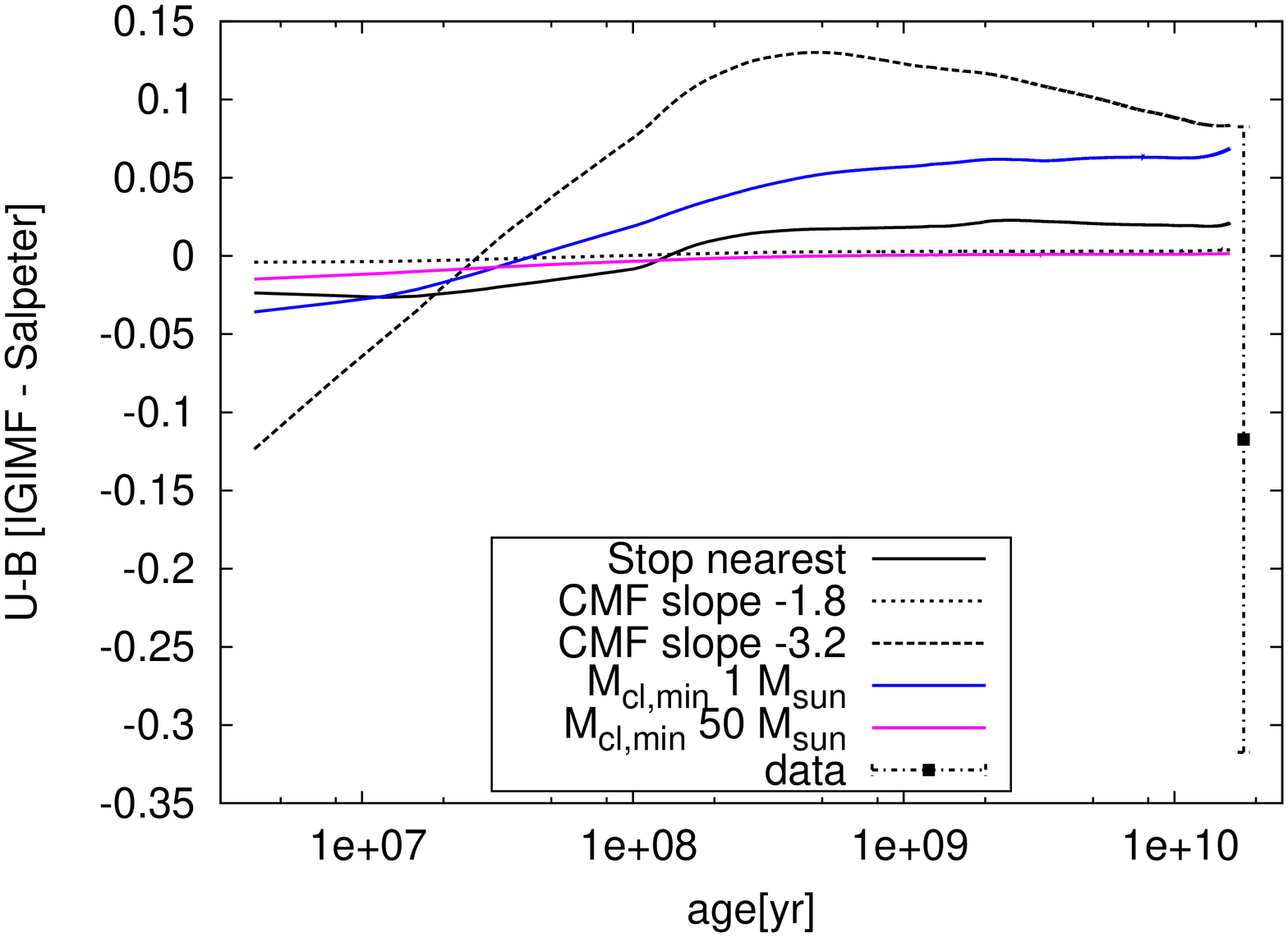} \\
\includegraphics[angle=270,width=0.45\linewidth]{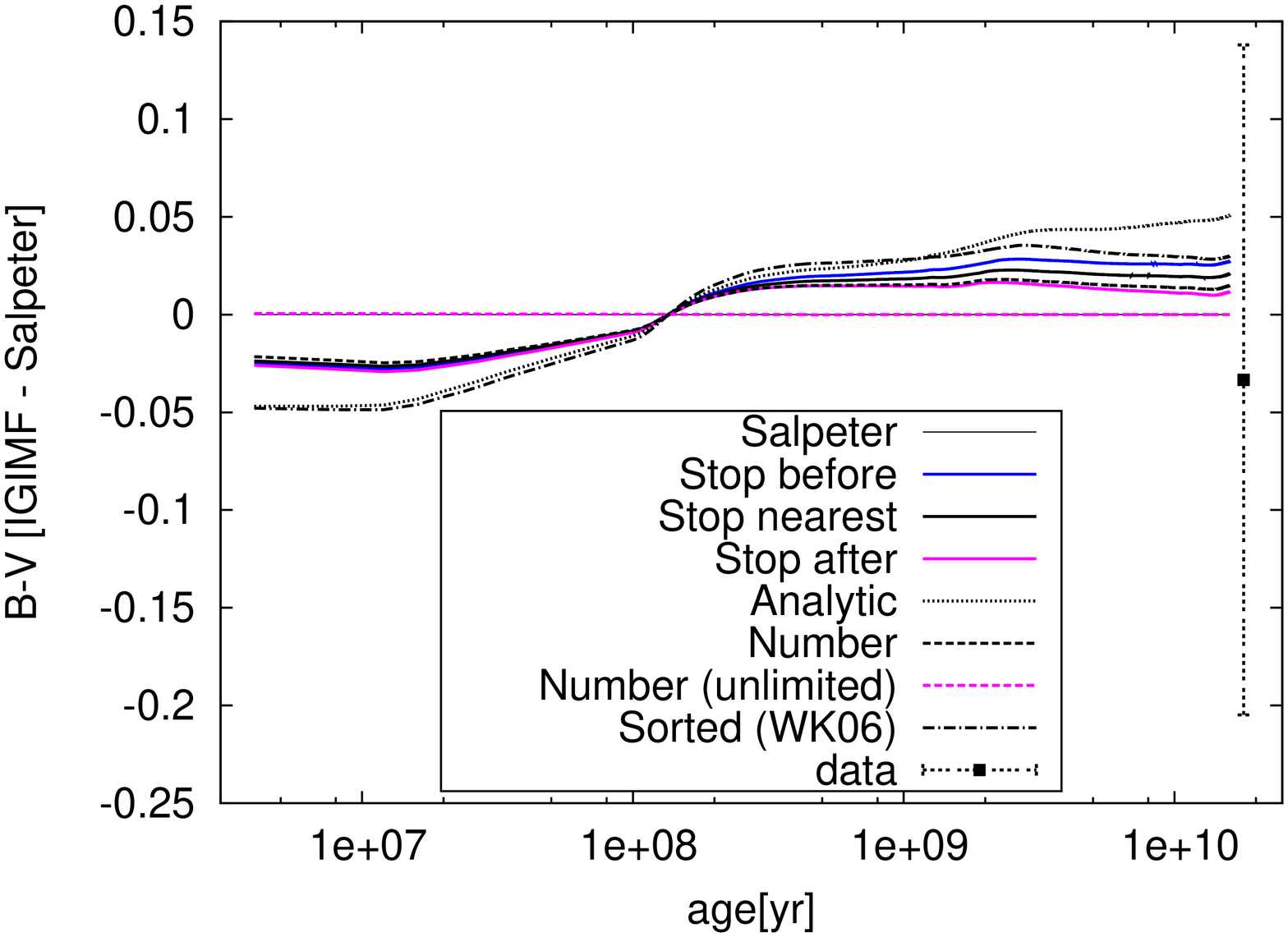} &&
\includegraphics[angle=270,width=0.45\linewidth]{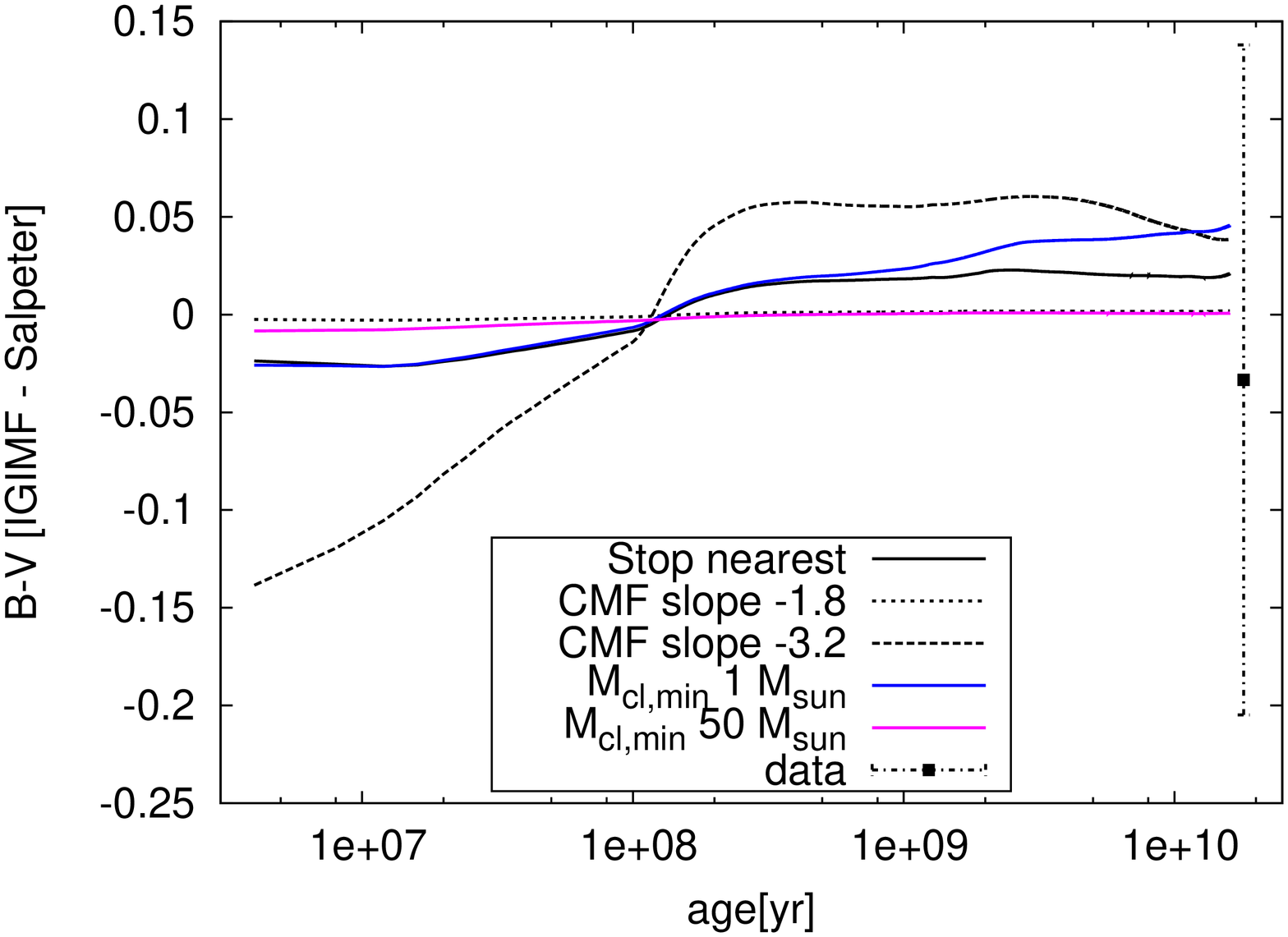} \\
\end{tabular}

\caption{The impact of various IGIMFs on the time evolution of
integrated Sd galaxy colours. Top row: U-B colour, bottom row: B-V
colour. Left column: IGIMFs for various sampling methods, right column:
IGIMFs for various CMF parameters. Shown are the differences between
models with different IGIMFs and the corresponding model with a
\citet{salpeter55} IMF. At the oldest ages, average colours (and their
standard deviation) for Sd galaxies from the HyperLeda database are
shown.} 

\label{fig:galev_sd} 
\end{figure*}

In Fig. \ref{fig:galev_sd} we compare our Sd galaxy models using
various IGIMFs with the standard model using the input
\citet{salpeter55} IMF (models for other Hubble types show very similar
behaviour). On the right side (i.e. plotted at old ages) we show
average colours and their standard deviation from data obtained from
the HyperLeda\footnote{http://leda.univ-lyon1.fr/} database
\citep{HYPERLEDA}, subdivided according to their morphological type. In
each of these plots, the intrinsic scatter within the morphological
type class well exceeds the deviations introduced by the different
IGIMFs. Therefore, we do not expect that IGIMF variations can be
constrained from integrated photometry of galaxies.

\subsection{Chemical enrichment in galaxies from different IGIMFs}

\begin{figure}
\centering
\includegraphics[angle=270,width=\columnwidth]{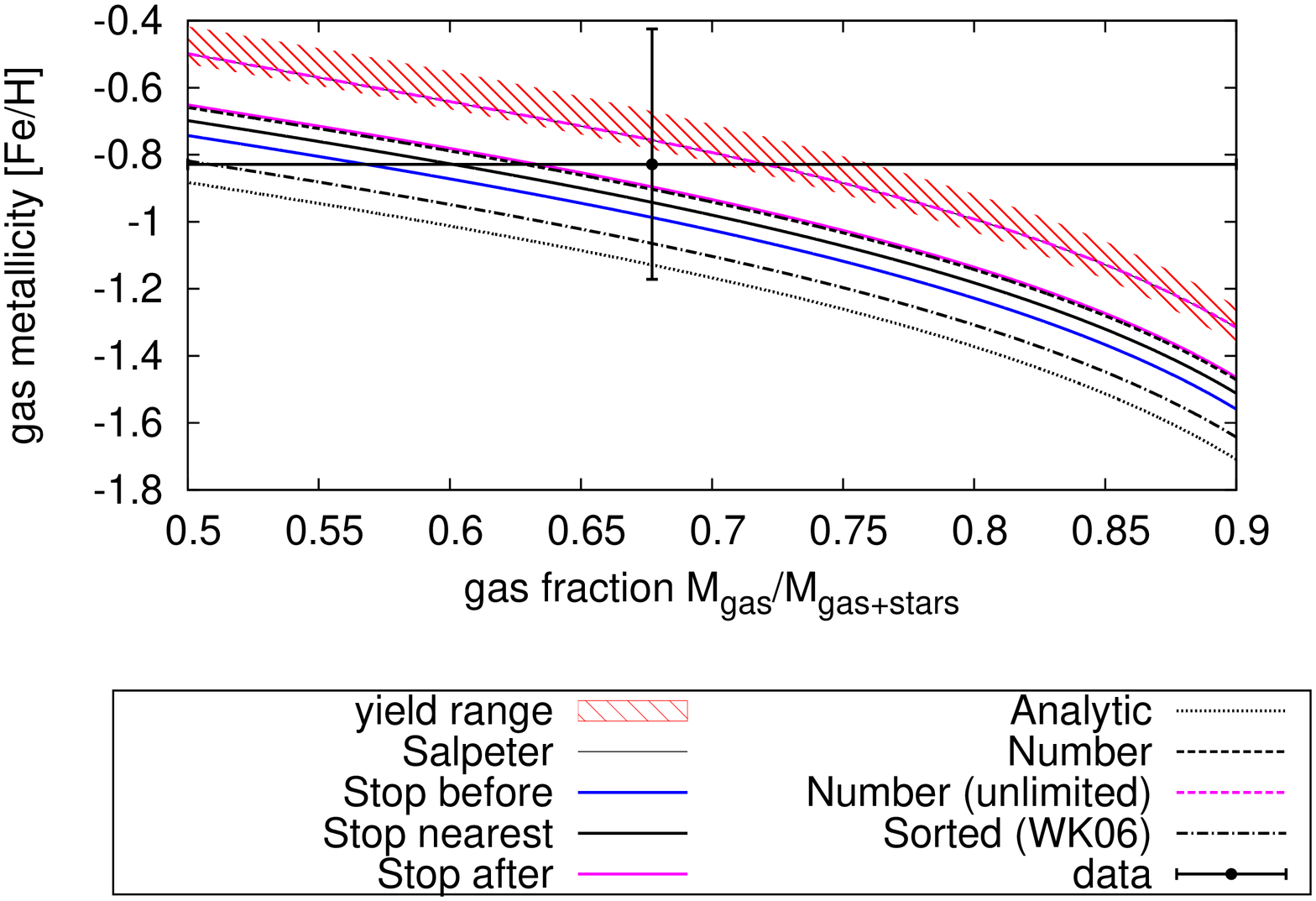}
\includegraphics[angle=270,width=\columnwidth]{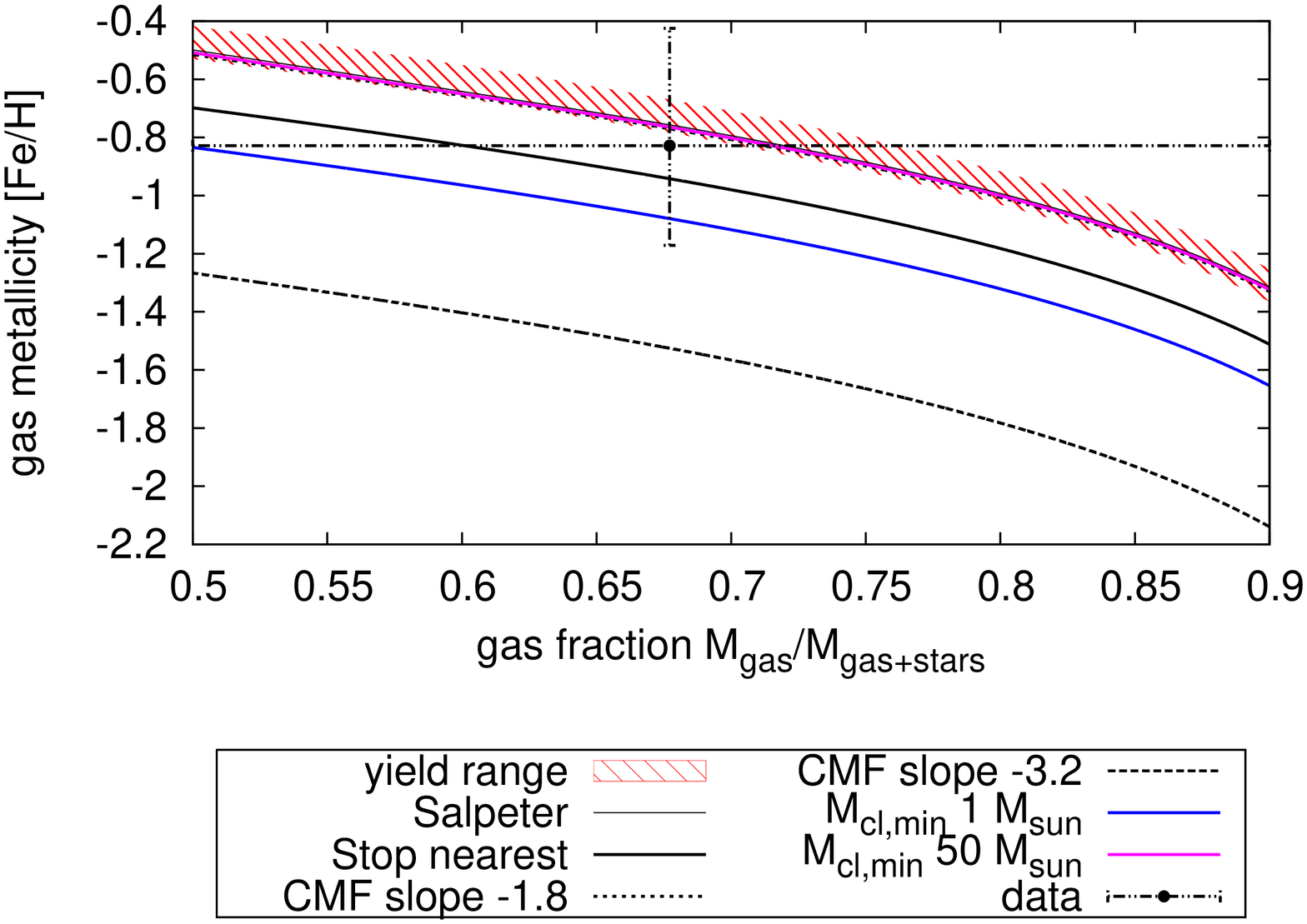}

\caption{The impact of various IGIMFs on the relation between available
gas mass and gas metallicity (e.g., chemical enrichment) for Sd
galaxies. Top panel: IGIMFs for various sampling methods. Bottom panel:
IGIMFs for various CMF parameters. The points mark estimates from
observations, see text for details. The red hashed area is the region
covered using  various individual metallicities, instead of the
``chemically consistent'' modeling, and a \citet{salpeter55} IMF. It
represents a worst-case uncertainty range.} 

\label{fig:galev_gas} 
\end{figure}

A more promising way might be the study of the gas properties in
galaxies. In Fig. \ref{fig:galev_gas} we show the relation between gas
fraction (i.e., the ratio between gas mass and gas + stellar mass) and
gas metallicity (we give all metallicities as $[$Fe/H$]$, assuming solar
abundance ratios and neglecting alpha-enhancement effects). Since the
majority of chemical enrichment originates in massive stars,
deficiencies of such stars due to IGIMF effects reflect directly in the
gas metallicity. The red hashed area is the region covered using 
various individual metallicities, instead of the ``chemically
consistent'' modeling, and represents a worst-case uncertainty range.
Consistent with this ``uncertainty region'' are 4 sets of models: the
input \citet{salpeter55} IMF models, the equivalent ``Number
(unlimited)'' models, and the models ``CMF slope = -1.8'' and ``M$_{\rm
cl,min}$ = 50 \msun''. This is in agreement with the little deviations
between the input \citet{salpeter55} IMF and these IGIMFs already seen
in Sect. \ref{sec:results}. The other models, using different IGIMFs are
clearly distinct from this ``uncertainty region'', with differences in
gas metallicity up to 1 dex, with various models offset by 0.2 -- 0.4
dex (corresponding to factors 1.5 -- 2.5).

To our best knowledge, there is no study which determines both gas
fractions and gas metallicities for a large sample of galaxies in a
consistent way. We therefore gather data on galactic gas masses from
\citet{huchtmeier89} and \citet{karachentsev99}, while for the gas
metallicities we consider catalogues by \citet{kewley05},
\citet{nagao06}, and \citet{izotov07}. These catalogues were not only
chosen for their (comparably) large sample sizes, but also for their
diversity in galaxy populations they address. Each of these samples has
its own intrinsic biases and limitations. \citet{huchtmeier89} and
\citet{nagao06} are rather literature compilation papers. The sample by
\citet{kewley05} intentionally contains galaxies of all Hubble types
with a wide range of properties, the \citet{karachentsev99} sample is
volume-limited, and \citet{izotov07} considers specifically
low-metallicity H{\sc ii} regions in nearby dwarf galaxies. We
supplement the catalogue information with data from the
HyperLeda$^1$ database
\citep{HYPERLEDA}, to have an as uniform as possible determination of
Hubble type and absolute luminosity for the sample galaxies. From these
data we estimate the average gas fractions and gas metallicities for the
galaxy samples. Where multiple observations for a given galaxy are
available, we include all of them individually, to access the
uncertainties more realistically, and to average out metallicity
gradients in a single galaxy. In Table \ref{tab:gasvalues} we present
the derived average values for the individual and the combined samples,
for 5 different galaxy types.

\begin{table*}
\caption{Average gas properties and integrated galaxy colours for various literature galaxy samples.}             
\label{tab:gasvalues}      
\centering                          
\begin{tabular}{c c c c c c c}        
\hline\hline                 
sample & galaxy type & \#galaxies & gas fraction & error gas fraction & &\\    
\hline                        
\citet{huchtmeier89} & E  & 46  & 0.36 & 0.16& &\\  	 
\citet{huchtmeier89} & Sa & 154 & 0.49 & 0.17& &\\
\citet{huchtmeier89} & Sb & 635 & 0.53 & 0.16& &\\
\citet{huchtmeier89} & Sc & 1284& 0.59 & 0.18& &\\
\citet{huchtmeier89} & Sd & 730 & 0.67 & 0.21& &\\ 
\citet{karachentsev99} & E  & 7  & 0.007 & 0.008& &\\  	 
\citet{karachentsev99} & Sa & 1  & 0.0003 & -& &\\
\citet{karachentsev99} & Sb & 6  & 0.015 & 0.015& &\\
\citet{karachentsev99} & Sc & 21 & 0.078 & 0.092& &\\
\citet{karachentsev99} & Sd & 45 & 0.19 & 0.14& &\\ 
combined & E  & 53   & 0.33 & 0.18 & &\\    
combined & Sa & 155  & 0.48 & 0.17 & &\\
combined & Sb & 641  & 0.52 & 0.16 & &\\
combined & Sc & 1305 & 0.59 & 0.18 & &\\
combined & Sd & 775  & 0.67 & 0.26 & &\\ 
\hline                        
sample & galaxy type & \#galaxies & $[$Fe/H$]$(gas) & error $[$Fe/H$]$(gas) & &\\    
\hline                        
\citet{kewley05} & E  & 9  & -0.09 & 0.18& &\\ 	  
\citet{kewley05} & Sa & 6  & -0.08 & 0.27& &\\
\citet{kewley05} & Sb & 18 & -0.09 & 0.27& &\\
\citet{kewley05} & Sc & 34 & -0.13 & 0.16& &\\
\citet{kewley05} & Sd & 18 & -0.42 & 0.32& &\\ 

\citet{nagao06} & E  & 3  & -0.74 & 0.22& &\\  	
\citet{nagao06} & Sa & 1  & -1.05 & -& &\\
\citet{nagao06} & Sb & 3  & -0.71 & 0.19& &\\
\citet{nagao06} & Sc & 5  & -1.07 & 0.34& &\\
\citet{nagao06} & Sd & 47 & -0.92 & 0.24& &\\ 

\citet{izotov07} & E  & 2  & -0.85 & 0.1& &\\    
\citet{izotov07} & Sa & 1  & -1.04 & -& &\\
\citet{izotov07} & Sb & 2  & -0.66 & 0.24& &\\
\citet{izotov07} & Sc & 8  & -0.84 & 0.30& &\\
\citet{izotov07} & Sd & 23 & -0.95 & 0.31& &\\ 

combined & E  & 14 & -0.34 & 0.39& &\\		
combined & Sa & 8  & -0.32 & 0.50& &\\
combined & Sb & 23 & -0.22 & 0.36& &\\
combined & Sc & 47 & -0.35 & 0.42& &\\
combined & Sd & 88 & -0.83 & 0.34& &\\ 
\hline                                   
sample & galaxy type & \#galaxies & U-B & error U-B & B-V & error B-V \\    
\hline                        
HyperLeda database &  E  & 547 &  0.36  & 0.21 & 0.83 & 0.13 \\
HyperLeda database &  Sa & 166 &  0.14  & 0.23 & 0.68 & 0.17 \\
HyperLeda database &  Sb & 329 &  0.02  & 0.19 & 0.61 & 0.16 \\
HyperLeda database &  Sc & 397 &  -0.10 & 0.15 & 0.50 & 0.13 \\
HyperLeda database &  Sd & 173 &  -0.23 & 0.20 & 0.39 & 0.17 \\
\hline                        
\end{tabular}
\end{table*}

From Table \ref{tab:gasvalues} one can easily see the non-homogeneity
of the samples. The multitude of biases and selection effects hampers a
straightforward comparison of these observational data with our models.
A dedicated survey of a large number of L$^*$ galaxies ({\sc galev}
attempts to model L$^*$ galaxies, and  therefore neglects galaxy
mass-dependent effects!) for the different Hubble types, both in terms
of gas fraction and in terms of gas metallicity, with a reliable
estimate of the galaxies' Hubble types, will be needed to provide
direct calibration values for our (and others) galaxy evolution models.

In Fig. \ref{fig:galev_gas}, we present gas properties for {\sc
galev} models of Sd galaxies, based on the various IGIMFs (equivalent
plots for galaxies of other  Hubble types appear very similar). We
include the data point corresponding to the combined data sets in  Table
\ref{tab:gasvalues}. As uncertainties we plot either the scatter for a
given property within the combined sample of Sd galaxies, or the
distance to the most deviating mean of any subsample, whichever is
larger.

Based on the large spread in the observed gas fractions it is hard 
to constrain IGIMF models with these data. Future and more homogeneous 
samples will be helpful, as the spread in observed metallicities is 
smaller than or at most comparable to the difference arising from 
different IGIMFs.

\section{Conclusions} \label{sec:conclusions}

We have conducted a suite of numerical experiments to investigate how
the steepening of the IGIMF depends on sampling method and the assumed
cluster mass function. Hereby, we extended the variations already
studied by WK06. We found that, unless the maximum occurring stellar
mass is not limited by the cluster mass or the minimum cluster mass is
higher than the maximum stellar mass, the IGIMF is always steeper at the
high-mass end than the input IMF, assuming that stars all form in clusters and that these cluster follow a power-law CMF which extends down to masses well below the upper mass limit for stars. If there are many stars formed in a non-clustered environment (see Eq.~\ref{eq:igimf_tot}) or the CMF does not go down to such low masses (or turns over and peaks at higher masses), IGIMF effects are much smaller, and possibly even become negligible. The amount of steepening and the mass
where deviations set in depend on the sampling method and the adopted
cluster mass function:

\begin{itemize}

\item The numerical method of sampling is important. 
While all random techniques result in the same high-mass slope (for 
constant CMF slope), the
onset of deviations occurs at different stellar masses, showing a slight 
steepening already at stellar masses below the lower cluster mass limit.

\item The slope of the cluster mass function, as well as its lower limit
are very important. 
The index of the cluster mass function sets the steepness of the IGIMF
at the high mass end: a steeper cluster mass function results in a
steeper IGIMF. Varying the CMF power-law index in the range $[-1.8, -2.2, 
-3.2]$ results in IGIMF slopes at the high mass end of approximately
$[-2.4, -2.6, -3.6]$.
The {\sl observationally ill-constrained} lower cluster mass limit sets the mass at which the steepening sets
in,  i.e., the IGIMF becomes much steeper from $m=M_{\textrm{\scriptsize
cl, min}}$. At slightly lower masses there is a very small deficiency
of stars as compared  to the input IMF. The magnitude of this
discrepancy depends on sampling method. Contrary to results by \citet{elmegreen06}, we do not find the $\beta=2$ CMF to be singular.

\end{itemize}

All sampling methods reproduce the input cluster mass functions well.
Even though some seem to steepen or shallow the CMF by construction,
the effects are marginal  and unobservable. The number of isolated
stars that should be formed according to  our method is very small (on
the order of 1 out of 10.000 clusters in our sample  consists of 1
star). The fraction of clusters consisting of only one O-star is even
an order of magnitude smaller. We also test the fraction of clusters
which are O-star dominated (clusters which contain an O-star which
represents at least half of the total cluster mass) to simulate
observational incompleteness, since a small underlying cluster might
stay unnoticed close to a bright O-star. This measure is rather
sensitive to the sampling method. For our default method we find about
0.56\% of such clusters, while for the ``sorted sampling'' by WK06 this
fraction is more than an order of magnitude lower. For sufficiently
large  samples of O-stars the O-star count could be a suitable tracer
of the IGIMF if the observed fraction of O-stars, delivered by surveys
like GAIA, is well understood.

Our default sampling results indicate that $\la$11\% of the 
O-stars in the Galaxy will be observed to be separate from any cluster 
environment, in nice agreement with results of \citet{dewit05}. The 
sorted sampling method of WK06 strongly underproduces this number.

However, current knowledge, both observationally and theoretically, of
the very formation processes of  (especially massive) stars in star
clusters (see e.g. high-mass star formation from high-mass cloud cores
\citep{krumholz05} vs competitive accretion
\citep{bonnell04}) prevents us from conclusions which
sampling method is favoured by nature. 

We conducted numerical experiments using  the {\sc galev} evolutionary
synthesis package, which self-consistently follows  the photometric and
chemical history of various idealized isolated galaxy models.  The
conclusions we draw on photometry and chemical enrichment resulting from
our  IGIMFs, as compared to the standard IMFs can be summarized as
follows:

\begin{enumerate}
\item Integrated photometry is likely not a good tracer of IGIMF
variations, since differences are smaller than the intrinsic
galaxy-to-galaxy scatter for a given morphological type.
\item Chemical enrichment is a better tracer as it is directly linked to
the number of massive stars, however, observations are rare and for
small sample sizes. Once the (systematic and random) uncertainties in 
determining gas mass fractions and metallicities are well understood, 
these quantities may be able to judge between several sampling methods 
(at least the ones with the most extreme deviations from the underlying IMF.).
\end{enumerate}

\noindent Future studies of galaxy evolution and chemical enrichment have to take
into account that the IGIMF is steeper than the normal IMF, as well as
the amount of uncertainty in the amount of steepening, as the details of
the sampling method nature chooses are poorly understood. Additional
uncertainties are introduced as the shape of the cluster mass function
is not well constrained at very low masses (i.e. cluster masses comparable to individual stellar masses), whereas the low mass end of the CMF is
the most important quantity in shaping the IGIMF. These differences between the
IMF and the IGIMF have pronounced implications for modeling galaxy
properties.

\begin{acknowledgements}

It is a pleasure to thank the referee, John Scalo, for a very helpful and 
thorough report that greatly improved the paper. We kindly thank Nate Bastian, 
Pavel Kroupa, Carsten Weidner and Simon Goodwin for
helpful comments on  an early version of the manuscript, as well
as Ralf Kotulla for reading and discussing details of the {\sc galev}
models. We acknowledge the usage of the HyperLeda database
(http://leda.univ-lyon1.fr).

\end{acknowledgements}

\bibliographystyle{aa} 

\bibliography{mybib}

\end{document}